\documentclass{PoS}

\PoS{PoS(LAT2005)233}

\title{ Towards a non-perturbative computation of the RGI strange
  quark mass with two dynamical flavors
\thanks{Preprint: HU-EP-05/46, SFB/CPP-05-51, DESY 05-172}}

\ShortTitle{Towards a non-perturbative computation of the RGI strange
  quark mass with two dynamical flavors}

\author{
\vspace{-0.5cm}
\vbox{
\epsfxsize=2.8 true cm
\epsfbox{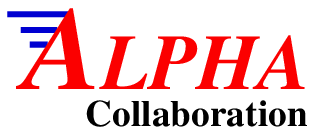}}
\vspace{0.3cm}
\speaker{Michele Della Morte}, Roland~Hoffmann, Francesco~Knechtli, Juri~Rolf,
	Ulli~Wolff\\
	Humboldt Universit\"at zu Berlin, Institut f\"ur Physik,\\
	Newtonstrasse~15, D-12489 Berlin, Germany\\
	E-mail: \email{\{dellamor,roland,knechtli,rolf,uwolff\}@physik.hu-berlin.de}}
\author{Rainer~Sommer \\ DESY, Platanenallee 6, 15738 Zeuthen, Germany \\
        E-mail: \email{rainer.sommer@desy.de}}
\author{Ines~Wetzorke \\ NIC/DESY , Platanenallee 6, 15738 Zeuthen, Germany \\
        E-mail: \email{ines.wetzorke@desy.de}}

\abstract{
The non-perturbative running of the quark mass in the Schr\"odinger
functional scheme is computed over a large energy range (covering scales
differing by two orders of magnitude). This allows to relate lattice
estimates of the running quark mass to the renormalization group
invariant mass. The result is used in a preliminary computation of the
strange quark mass in the theory with two flavors of non-perturbatively
improved Wilson quarks. A more detailed discussion of the calculation can
be found in~\cite{ms2}.
}
\FullConference{XXIIIrd International Symposium on Lattice Field Theory\\
		 25-30 July 2005\\
		 Trinity College, Dublin, Ireland}
\usepackage{amssymb,amsmath}
\usepackage{epsfig}
\usepackage{psfrag}
\newcommand\ba{\begin{array}}
\newcommand\ea{\end{array}}
\newcommand{\be}{\begin{equation}}
\newcommand{\ee}{\end{equation}}
\newcommand{\bea}{\begin{equation}\begin{array}}
\newcommand{\eea}{\end{array}\end{equation}}
\newcommand{\bdm}{\begin{displaymath}}
\newcommand{\edm}{\end{displaymath}}
\newcommand{\rba}{\begin{array}}
\newcommand{\rea}{\end{array}}
\newcommand{\bi}{\begin{itemize}}
\newcommand{\ei}{\end{itemize}}
\newcommand{\bver}{\begin{verbatim}}
\newcommand{\everbatim}{\end{verbatim}}
\newcommand{\Ffrac}[2]{\frac{\displaystyle{#1}}{\displaystyle{#2}}}
\def\mbar{\overline{m}}
\def\ZP{Z_{\rm P}}
\def\sp{\sigma_{\rm P}}

\def\ZA{Z_{\rm A}}
\def\dd{{\rm d}}
\def\gbar{\bar{g}}
\def\Lmax{L_{\rm max}}

\begin{document}
\section{Introduction}
\vspace{-0.15cm}
Quark masses are parameters of QCD. They can be
computed non-perturbatively on the lattice once an hadronic input is
given.  Sticking
to the renormalization group invariant (RGI) definition of the quark 
mass, lattice results for the strange quark mass with
2~\cite{AliKhan:2001tx, Aoki:2002uc, Gockeler:2004rp} and
2+1~\cite{Ishikawa:2004xq, Aubin:2004ck}
 dynamical flavors differ among each other by more than 40 MeV. Such a
 spread can be ascribed to various systematics, mainly cutoff
 effects and the renormalization procedure, which in many cases relies
 on perturbation theory only. 
For the result we will present here the  renormalization is performed in a
fully non-perturbative way, while the lattice spacing is still varied
in a small range only (roughly 0.1 to 0.07 fm). 

We write the relation
among the bare current (PCAC) quark mass $m_i$ and the RGI mass $M_i$ as
\be
M_i = Z_{\rm M}(g_0) \, m_i(g_0) \,, \label{e:zM}
\ee
where $i$ has to be interpreted as a flavor index. The factor $Z_{\rm  M}$
 can be split into two parts, the first one connecting the bare
mass to $\mbar_i(\mu)$, the renormalized one (in the Schr\"odinger
functional scheme) at a given scale $\mu$
\be
\mbar_i(\mu)={\ZA(g_0) \over \ZP(g_0,a\mu)}\, m_i(g_0) \,,
\label{rmass}
\ee
and the second one $M/\mbar(\mu)$, which relates two renormalized masses and is
therefore universal. In mass independent schemes (like the one we are
going to describe) this second factor is also flavor
independent~\cite{Leutwyler:1996qg}. Its computation is the main result
we report about here. 

The renormalization group equations for the
couplings are decoupled in mass independent schemes and read
\begin{equation}
\mu\frac{\dd\gbar}{\dd\mu}=\beta(\gbar)\quad,\quad 
\mu\frac{\dd\mbar_i}{\dd\mu}=\tau(\gbar)\mbar_i \,.
\label{RGIeq}
\end{equation}
with $\gbar(\mu)$ the renormalized coupling at the scale $\mu$.
The $\tau$ and $\beta$ functions can be expanded in perturbation theory
\be
\beta(\gbar) 
 _{\mbox{$\stackrel{\displaystyle\sim}{\scriptstyle \gbar\rightarrow0}$}} 
 -\gbar^3\{b_0+b_1\gbar^2+b_2\gbar^4+...\}  \quad ,\quad
 \tau(\gbar) 
 _{\mbox{$\stackrel{\displaystyle\sim}{\scriptstyle \gbar\rightarrow0}$}} 
 -\gbar^2\{d_0+d_1\gbar^2+...\} \,, \label{taufunc}
\ee
and the exact (i.e. valid beyond perturbation theory) solutions of
eqs.~(\ref{RGIeq}) can be written
\begin{eqnarray}
 \Lambda & = & \mu(b_0\gbar^2)^{-b_1/2b_0^2}
 {\rm e}^{-1/(2b_0\gbar^2)}
 \exp\left\{-\int_0^{\gbar}\dd x\left[
 \Ffrac{1}{\beta(x)}+\Ffrac{1}{b_0x^3}-\Ffrac{b_1}{b_0^2x}\right]\right\}
 \,,\label{LAMBDA} \\
 M_i & = & \mbar_i(2b_0\gbar^2)^{-d_0/2b_0}
 \exp\left\{-\int_0^{\gbar}\dd x\left[
 \Ffrac{\tau(x)}{\beta(x)}-\Ffrac{d_0}{b_0x}\right]\right\}
 \,,\label{MRGI}
\end{eqnarray}
where the RGI parameters $\Lambda$ and $M$ appear as integration constants.
At the level of RGI parameters the connections among different schemes
can be given in a  simple and exact way. We thus regard them as
the fundamental parameters of QCD.
\section{Running of the quark mass in the SF scheme}
In terms of Schr\"odinger functional (SF) correlation functions the
bare PCAC mass  reads
\be
 m(g_0,\kappa) = 
               \left. \frac{\frac{1}{2}(\partial_0^{\ast}+\partial_0) f_{\rm A}(x_0)
                +c_{\rm A} a\partial_0^{\ast}\partial_0 f_{\rm P}(x_0)}
               {2f_{\rm P}(x_0)}\right|_{x_0=T/2}\; , \label{PCACmass}
\ee
where $\kappa$ is the hopping parameter. We use the plaquette gauge action
and non-perturbatively
O($a$) improved Wilson fermions~\cite{csw2}. The
improvement coefficient $c_{\rm A}$ is also set to its
non-perturbative value from~\cite{cA2}. For the definition of the SF
with Wilson fermions we refer to~\cite{SF} and to~\cite{nf0} for any
unexplained notation. Here we only recall that the correlation
functions $f_{\rm A}$ and $f_{\rm P}$ involve matrix elements of the
axial current and the pseudoscalar density operators, respectively.
According to eq.~(\ref{rmass}) the running quark mass at the scale $\mu$
is obtained by multiplying the mass in eq.~(\ref{PCACmass}) by the ratio 
$Z_{\rm A}(g_0)/Z_{\rm P}(g_0, a\mu)$. The factor $Z_{\rm A}$ is scale
independent and can be fixed by Ward identities~\cite{ZA2}. All the
scale dependence in the running quark mass comes from  $Z_{\rm P}$. 
In the SF it can be defined as
\be
Z_{\rm P} (g_0,L/a)= c {{\sqrt{3f_1}}\over f_{\rm P}(L/2)}\;, \quad
m=0, \;\; \theta=0.5, \;\; T=2L\;,
\label{ZP}
\ee
where the boundary to boundary correlation function $f_1$ takes care
of the renormalization of the boundary quark fields and the constant
$c$ ensures $Z_{\rm P}=1$ at tree level. Eq.~(\ref{ZP})
makes clear that in the finite volume scheme we are using, the
normalization scale $\mu$ is identified with the (inverse) extent of
the system, which, in turn, is uniquely fixed (up to cutoff effects)
by the value of the running coupling $\bar{g}^2(L)$.
 The non perturbative running of the quark mass is
described by the step scaling function $\sigma_{\rm P}(u)$
\be
\sigma_{\rm P}(u)={{\mbar(\mu)}\over{\mbar(\mu/2)}}=\lim_{a \rightarrow 0} 
\left.{{Z_{\rm P}(g_0,2L/a)}\over{Z_{\rm
      P}(g_0,L/a)}}\right|_{\gbar^2(L)=u} \;,
\ee
which can be viewed as an integrated form of the $\tau$ function.
Introducing also the step scaling function for the coupling 
$\sigma(u)=\gbar^2(L)$ (computed in~\cite{nf2coup}) and solving the following coupled
recursions 
\begin{eqnarray}
\left\{\ba{l} u_0=\gbar^2(\Lmax)=4.61 \\ \sigma(u_{k+1})=u_k \ea\right.
& \Rightarrow & u_k=\gbar^2(L_k) \,,\quad L_k = 2^{-k}\Lmax \,, \label{ucoeff} \\
\left\{\ba{l} w_0=1 \\ w_k=\left[\prod_{i=1}^k\sp(u_i)\right]^{-1} \ea\right.
& \Rightarrow & w_k=\frac{\mbar(1/\Lmax)}{\mbar(1/L_k)} \,, \label{wcoeff}
\end{eqnarray}
starting from the low energy scale $1/\Lmax$ defined by
 $u_0 \,=\,\gbar^2(\Lmax)  =  4.61$
to the higher scales $1/L_k$, $k=0,1,\ldots,8$ (with $L_0\equiv\Lmax$),
one can obtain the ratio $\frac{\mbar(1/\Lmax)}{\mbar(1/L_k)}$. For
$k\geq 6$ contact can be made with perturbation theory, in such a way
that the RGI parameters can be computed from
eqs.~(\ref{LAMBDA},~\ref{MRGI}) by using the
2-loop $\tau$ function and 3-loop $\beta$ function.

\begin{figure}[h]
\vspace{-1.92cm} 
\begin{center}
     \includegraphics[width=9.82cm]{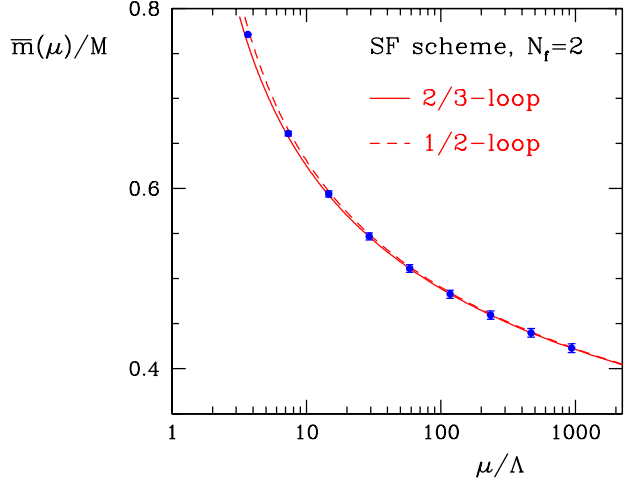}
\end{center}
 \vspace{-2.1cm}
 \caption{The non--perturbative running of $\mbar$.
 2/3--loop refers to the 2--loop
 $\tau$-function and 3--loop $\beta$--function, analogously 1/2--loop.}
 \label{f_run}
\vspace{-0.2cm}
\end{figure}

In practice we calculated $\sigma_{\rm P}(u)$ at 6 different couplings in
the range $u=3.33 \dots 0.98$, extrapolating to the continuum limit
lattice results from resolutions $L/a=6,$ 8 and 12 (see~\cite{ms2} for
details about the extrapolations). The change in scale by a factor 2 
needed for $\sigma$ and $\sigma_{\rm P}$ is ``easily'' implemented in
the SF scheme as it amounts to doubling $L/a$ at fixed bare parameters
(in other words we used lattices with $L/a$ up to 24). We parameterize
the results by the ansatz
\be
\sigma_{\rm P}(u)= 1-\ln(2)d_0 u+ p_1u^2+p_2 u^3 \;, 
\ee
with free coefficients $p_1$ and $p_2$. Inserting it in the
recursion eq.~(\ref{wcoeff}), for the case $k=6$, we obtain
\be
\frac{M}{\mbar(\mu)} =  1.297(16) \quad \mbox{at} \quad
 \mu \,=\, 1/\Lmax \,. \label{MombarLmax}
\ee
Using the result for $\Lmax\Lambda$ from~\cite{nf2coup}, we plot in figure~\ref{f_run}
the running of the quark mass versus $\mu/\Lambda$. In the plot we
only include the errors from the coefficients $w_k$; the uncertainties
on $\Lambda$ and $M$ would simply amount to a shift of the axes.
The figure shows that in this case perturbation theory works surprisingly well
down to rather small energies. This property is anyway specific of the
quantity and the scheme considered here. As an example, the
non-perturbative $\beta$ function in the SF scheme shows larger
deviations 
from perturbation theory at the lower end of energies reached here~\cite{nf2coup}.

\vspace{-0.13cm} 
\section{The strange quark mass}
\begin{figure}[h]
 \begin{center}
     \includegraphics[width=9.95cm]{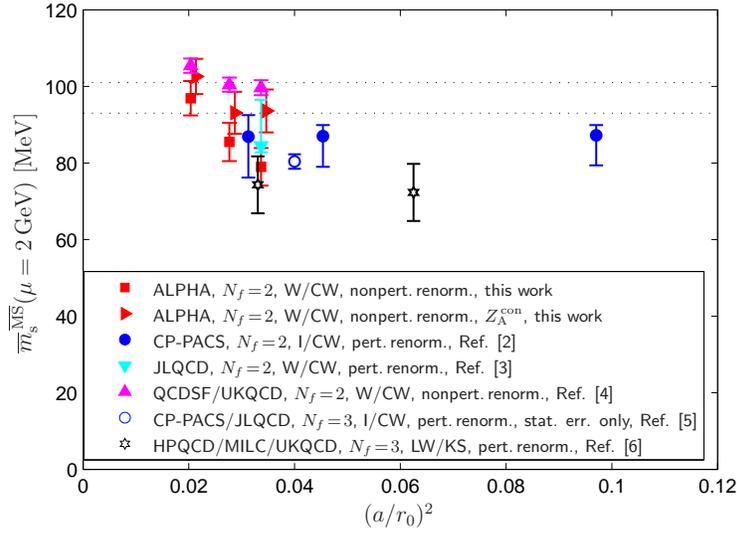}
 \end{center}
 \vspace{-0.6cm}
 \caption{Summary of the strange quark mass data from lattice simulations.
 In the legend the discretizations used are indicated in the form
 gauge action/fermion action. The dictionary reads:
 W: Wilson gauge action; I: Iwasaki gauge action;
 LW: 1--loop tadpole improved L{\"u}scher--Weisz gauge action;
 CW: Wilson--clover fermion action; KS: Asqtad staggered fermion action.
 The dotted lines represent the quenched result \cite{msnf0}. $r_0=0.5$
 fm has been used for all the datasets, whereas in~\cite{ms2}
 $r_0=0.467$ fm was used for the data from~\cite{Gockeler:2004rp}, as
 suggested thereby.}
 \label{f_allms}
\end{figure}

\vspace{-0.13cm}
In order to apply the described procedure to the computation of the
strange quark mass we first need to calculate the factor
$Z_{\rm A}(g_0)/Z_{\rm P}(g_0,\Lmax/a)$ 
for the bare couplings (i.e. lattice spacings) used in large
volume simulations. As we are going to use results
from~\cite{Gockeler:2004rp}, the relevant set of $\beta$ values is
5.2, 5.29 and 5.4. Since this region is covered by the result for $Z_{\rm A}(g_0)$
(and for $Z_{\rm A}^{\rm con}$, see
figure~\ref{f_allms})~in~\cite{ZA2},
 it remains to compute $Z_{\rm P}(g_0,\Lmax/a)$. At $\beta=5.2$  
the value is directly obtained by a simulation at $L/a=6$ while at the
other couplings we had to interpolate the results from different $L/a$
(the appropriate one wouldn't have been an integer). The numbers are
collected in table~\ref{t_ZM}. Clearly now $Z_{\rm P}$ and $Z_{\rm M}$
refer to the chosen discretization, whereas the ratio in
eq.~(\ref{MombarLmax}) is universal.
\vspace{-0.3cm}
\begin{table}[h]
 \centering
  \begin{tabular}{lll}
   \hline
   \multicolumn{1}{c}{$\beta$}  &
   \multicolumn{1}{c}{$\ZP$} &
   \multicolumn{1}{c}{$Z_{\rm M}$} \\[.5ex]
   \hline
5.20 & 0.47876(47) & 1.935(33)(24) \\
5.29 & 0.4936(34)  & 1.979(25)(24) \\
5.40 & 0.4974(33)  & 2.001(29)(25) \\
   \hline
  \end{tabular}
 \caption{Results for $\ZP$ and $Z_{\rm M}$ for three coupling values.}
 \label{t_ZM}
\vspace{-0.25cm}
\end{table}

As we work with two degenerate flavors, instead of the strange quark
mass what we actually compute is $M_{\rm ref}$ associated with a kaon
made from two degenerate quarks. We use data for $r_0(\kappa)$ and
$m_{\rm PS}(\kappa)$ from~\cite{Gockeler:2004rp}. After extrapolating 
$r_0(\kappa)$ to the chiral limit (i.e to $\kappa=\kappa_c$) we fit
the product $r_0(\kappa_c)m_{\rm PS}(\kappa)$ as a function of $\kappa$
in order to determine $\kappa_{\rm ref}$ defined such that
\be
(r_0(\kappa_c)m_{\rm PS}(\kappa_{\rm ref}))^2=(r_0m_{\rm K})^2=1.5736
\; ,
\ee
where we have used $r_0=0.5\;$ fm~\cite{r0}. Finally we compute the
PCAC masses at the bare parameters $\beta,\kappa_{\rm ref}(\beta)$ with
$\beta=$5.2, 5.29 and 5.4  in volumes of approximately 1.5 fm. This
reduces cutoff effects in the PCAC mass due to finite
size~\cite{large}. The result for the RGI quantity is
\be
M_{\rm ref} =72(3)(13) \; {\rm MeV} \; ,
\label{mref_res}
\ee
the first error is statistical, the second is our estimate of cutoff
effects obtained by comparing the result at the finest lattice
spacing with that at the coarsest one. The number in
eq.~(\ref{mref_res}) is consistent with the quenched estimate of the
same quantity~\cite{nf0}. Assuming a weak dependence on $N_{\rm f}$ 
we  re-interpret our result as an estimate in the 3-flavor
theory (this assumption of course has to be checked in the future),
 and relate $M_{\rm ref}$ to $M_{\rm s}$ through the lowest order
chiral perturbation theory formula~\cite{galeut}
\be
 m_{\rm K}^2 \,=\,
 \frac{1}{2}(m_{{\rm K}^+}^2 + m_{{\rm K}^0}^2) =  (\hat{M}+M_{\rm
   s}) \, B_{\rm RGI}
 \label{mKsq} \,,
\ee
where $\hat{M}=1/2(M_{\rm u}+M_{\rm d})$, and
$B_{\rm RGI}$ is a constant of the chiral Lagrangian. For degenerate
quarks of mass $M_{\rm ref}$ eq.~(\ref{mKsq}) reads $m_{\rm K}^2= 2
M_{\rm ref} B_{\rm RGI}$, which implies $M_{\rm ref}=(\hat{M}+M_{\rm
  s})/2$. Using the relation $M_{\rm s}/\hat{M}=24.4(1.5)$ \cite{Leutwyler:1996qg}
we obtain 
\be
M_{\rm s} \approx 48/25 M_{\rm ref} \;\; \Rightarrow \;\; M_{\rm
  s}=138(5)(26) \; {\rm MeV} \;.
\ee
Higher order contributions from chiral perturbation theory are
expected to be around $10\%$ and thus below the accuracy we have
reached here. Eventually the result can be converted to the
$\overline{\rm MS}$
scheme at the scale 2 GeV by employing the 4-loop $\tau$ and $\beta$
functions~\cite{MSPT} for $N_{\rm f}=2$. This yields $\overline{m}_{\rm
    s}^{\overline{\rm MS}}(2\; {\rm GeV})=97(22)$ MeV. 

We give our conclusions discussing the collection of results in 
figure~\ref{f_allms}. By comparing different O($a$) improved
regularizations we see that cutoff effects on the strange quark
mass at a lattice spacing of $0.1$ fm are around 20\%. It is more
difficult to assess the $N_{\rm f}$ dependence mainly because only
perturbative renormalization has been used in the 3-flavor case.
On the other hand the comparison of $N_{\rm f}=2$ determinations
indicates that the use of perturbation theory for the renormalization
constants underestimates the quark mass.

{\bf Acknowledgement.}
We  thank NIC/DESY  for allocating computer time on the APEmille machines
and the APE group for their support.
This work has been supported by the
SFB Transregio 9  and by
the Deutsche Forschungsgemeinschaft 
in the Graduiertenkolleg GK 271 as well as by the
European Community's Human Potential Programme under contract HPRN-CT-2000-00145.

\vspace{-0.3cm}
%
\bibliographystyle{JHEP}
\providecommand{\href}[2]{#2}\begingroup\raggedright\endgroup

\end{document}